\documentclass[12pt]{article}
\usepackage{bbm}

\input epsf
\usepackage{amssymb}
\usepackage[dvips]{graphicx}
\usepackage{amscd}
\usepackage{mathrsfs}
\usepackage{amsmath}

\newcommand{\be}{\begin{equation}}
\newcommand{\ee}{\end{equation}}
\newcommand{\beqa}{\begin{eqnarray}}
\newcommand{\eeqa}{\end{eqnarray}}
\newcommand{\nn}{\nonumber}

\begin{document}
\linespread{1.5} \flushbottom

\begin{center}
{\Large\bf{ Coulomb problem in non-commutative quantum mechanics -
Exact solution}} \vskip1cm
V. G\' alikov\' a and P. Pre\v{s}najder\\
{\it Faculty of Mathematics, Physics and Informatics,\\
Comenius University Bratislava, Slovakia}\\ 
presnajder@fmph.uniba.sk

\end{center}

\begin{abstract}
We investigate consequences of space non-commutativity in quantum
mechanics of the hydrogen atom. We introduce rotationally
invariant noncommutative space $\hat{\bf R}^3_0$ - an analog of
the hydrogen atom ($H$-atom) configuration space ${\bf R}^3_0\,=\,
{\bf R}^3\setminus \{0\}$. The space $\hat{\bf R}^3_0$ is
generated by noncommutative coordinates realized as operators in
an auxiliary (Fock) space ${\cal F}$. We introduce the Hilbert
space $\hat{\cal{H}}$ of wave functions $\hat{\psi}$ formed by
properly weighted Hilbert-Schmidt operators in ${\cal F}$.
Finally, we define an analog of the $H$-atom Hamiltonian in $\hat{\bf
R}^3_0$ and explicitly determine the bound state energies
$E^\lambda_n$ and the corresponding eigenstates
$\hat{\psi}^\lambda_{njm}$. The Coulomb scattering problem in
$\hat{\bf R}^3_0$ is under study.
\end{abstract}

\newpage

\section{Introduction}

Basic ideas of non-commutative geometry have been developed in \cite{Con}
and, in a form of matrix geometry, in \cite{Mad1}. The main applications
have been considered

$\bullet $ in the area of quantum quantum field theory in order to understand,
or even to remove, UV singularities, and

$\bullet $ eventually to formulate a proper base for the quantum gravity.

The analysis performed in \cite{DFR} led to the conclusion that quantum
vacuum fluctuations and Einstein gravity could create  (micro)black holes
which prevent localization of space-time points. Mathematically this
requires non-commutative coordinates $x^\mu$ in space-time satisfying
specific uncertainty relations. The simplest set of operators $\hat{x}^\mu$
representing $x^\mu$ in an auxiliary Hilbert space should satisfy
Heisenberg-Moyal commutation relations
\be\label{DFR1} [\hat{x}^\mu,\hat{x}^\nu]\ =\ i\,\theta^{\mu\nu},\ \
\mu,\nu = 0,1,2,3\,,\ee
where $\theta^{\mu\nu}$ are given numerical constants that specify the
non-commutativity of the space-time in question.

Later in \cite{Jab} it was shown that field theories in NC spaces
with (\ref{DFR1}) can emerge as effective low energy limits of
string theories. These results supported a vivid development of
non-commutative QFT. However, such models contain various
unpleasant and unwanted features. The divergences are not removed, 
on the contrary, UV-IR mixing appears, \cite{UVIR}. The Lorentz 
invariance is broken down to $ SO(2)\times SO(1,1)$, but even this is 
sufficient to prove the classical CPT and Spin-statisics theorems, 
\cite{CPT}. This was not accidental and led to the twisted Poincar\'e 
reinterpretation of NC space-time symmetries, \cite{TwP}. 

However, it could be interesting to reverse the approach. Not to
use the NC geometry to improve the foundation of QFT,  what is a
very complicated task, but to test the effect of non-commutativity
of the space on the deformation of the well-defined quantum
mechanics (QM):

$\bullet $ Various QM systems have been investigated in 3D space
with Heisen\-berg-Moyal commutation relations
$[\hat{x}_i,\hat{x}_j] = i \theta^{ij}$, $i,j = 1,2,3$, e.g.
harmonic oscillator, Aharonov-Bohm effect, Coulomb problem, see
\cite{AB}, \cite{CP}. However, in such 3D NC space the rotational
symmetry is violated and there are systems, such as $H$-atom, that
are tightly related to the rotational symmetry.

$\bullet $ The rotational symmetry survives in 2D Heisenberg-Moyal
space with NC coordinates $\hat{x}_1,\hat{x}_2$ satisfying the
${\cal F}$ commutation relations $[\hat{x}_1,\hat{x}_2] = i
\theta$ in an auxiliary Hilbert space. In \cite{Scholtz} a planar
spherical well was described in detail:

(i) First, the Hilbert space ${\cal H}$ of operator wave functions
$\hat{\psi } = \psi (\hat{x}_1,\hat{x}_2)$ was defined;

(ii) Further, the Hamiltonian was defined as an operator acting in
${\cal H}$. It was nice to see how the persisted rotational
symmetry helps to solve exactly the problem in question.

The presented list of references is incomplete and we apologize
for that. We restricted ourselves to those which initiated
progress or are close to our approach.

Our aim is to extend this scheme to the QM problems with
rotationally symmetric potentials $V(r)$ in the configuration
space $R^3_0\,\equiv\,R^3 \setminus\{0\}$. We restrict ourselves
to the Coulomb potential which, in the usual (commutative)
setting, is a solution of the Poisson equation finite at infinity:
\be\label{coul0} \Delta V(r)\,=\,0\ \ \Rightarrow\ \ V(r)\ =\
-\,\frac{q}{r}\ +\ q_0 \,. \ee
For $H$-atom $q$, in a Gaussian system of units, is a square of 
electric charge $e^2$, and we put the inessential constant 
$q_0 = 0$. In this case we are dealing with Schr\" odinger 
equation
\be\label{Schr0} - \frac{\hbar^2}{2m}\,\Delta \psi({\bf
x})\,-\,\frac{e^2}{r} \psi({\bf x})\,=\,E \psi({\bf x}),\ \ r =
|{\bf x}| > 0 \ee
in the Hilbert space ${\cal H}$ specified by the norm
\be\label{norm0} \|\psi\|^2_0\,=\,\int\,d^3{\bf x}\ |\psi({\bf
x})|^2 \,. \ee
Expressing the wave function as
\be\label{psi0} \psi({\bf x})\,=\,R_j(r)\,H_{jm}({\bf x}),\ \
H_{jm}({\bf x}) \,\sim\,r^j\,Y_{jm}(\vartheta,\varphi)\,, \ee
and putting $\alpha = 2m e^2/\hbar^2$ and $\kappa = \sqrt{-2m E} /
\hbar$, we obtain the radial Schr\" odin\-ger equation:
\be\label{rSchr0}  r\,R^{\prime\prime}_j(r) + 2(j+1) R^\prime_j(r)
+ \alpha R_j(r)\,=\,\kappa^2 r\,R_j(r)\,. \ee
Its solution is
\be\label{rad0} R_j(r)\,=\,e^{-\kappa r}\,F\left(j+1-
\frac{\alpha}{2\kappa},\,2j+2;\,2\kappa r\right)\,, \ee
where $F(a,c;x)$ is the confluent hypergeometric function.

For bound states $E<0$, i.e. real-valued $\kappa$, the solution should
have a finite norm in ${\cal H}_0$. This is the case when the first
argument of the degenerated hypergeometric function is zero or a
negative integer, and this determines the discrete energy
eigenvalues:
\be\label{ener0}
\frac{\alpha}{2\kappa_n}\,=\,n\,=\,j+1,\,j+2,\,\dots\ \ \
\Rightarrow\ \ \ E_n\,=\,-\frac{\hbar^2}{2m}\,\kappa^2_n\,=\,
-\frac{m\,e^4}{2\hbar^2 n^2}\ . \ee
The hypergeometric function then reduces to a polynomial of degree
$n-j-1$ in the variable $x\,=\,2 r \kappa$:
\be\label{deghf}F(j+1-n,2j+2;x)\,=\,\sum_{k=0}^{n-j-1} c^k_{nj}
\frac{ (-x)^k}{k!}\,,
\ \ c^k_{nj}\,=\,\frac{(n-1)!(2j+1)!}{(n-1-k)!(2j+1+k)}\ .\ee

In this paper we extend the QM solution
(\ref{coul0})-(\ref{ener0}) of the Coulomb problem in $R^3_0$ to
the non-commutative rotationally invariant space. In Section 2 we
define $\hat{R}^3_0$ ,  a rotationally invariant NC generalization of
the configuration space $R^3_0$, we introduce the generators of
rotations and the NC analog of their eigenfunctions. In Section 3
we define Hilbert space $\hat{\cal H}$ - the NC analog of ${\cal
H}$, and we introduce the NC analog of the Coulomb problem Hamiltonian
acting in $\hat{\cal H}$. In Section 4 we exactly solve the
corresponding NC analog of the Schr\"odinger equation. Last Section
5 contains conclusions and perspectives.

\section{The noncommutative space $\boldsymbol{\hat{R}^3_0}$}

In this section we define the noncommutative space ${\bf
\hat{R}}^3_0$, possessing full rotational invariance, as a sequence
of fuzzy spheres introduced, in various contexts, in \cite{Ber}.
Different fuzzy spheres are related in such a way that at large
distances we recover space ${\bf R}^3_0$ with the usual flat geometry.
A similar construction of a 3D noncommutative space, as a sequence
of fuzzy spheres, was proposed in \cite{Jab1}. However, various
fuzzy spheres are related to each other differently (not leading
to the flat 3D geometry at large distances).

We realize the noncommutative coordinates in ${\bf\hat{R}}^3_0$ in
terms of 2 pairs of boson annihilation and creation operators
$\hat{a}_\alpha$, $\hat{a}^\dagger_\alpha$, $\alpha\,=\,1,2$,
satisfying the following commutation relations, see \cite{GP}:
\be [\hat{a}_\alpha,\hat{a}^\dagger_\beta]\,=\,\delta_{\alpha
\beta },\ \
[\hat{a}_\alpha,\hat{a}_\beta]\,=\,[\hat{a}^\dagger_\alpha,
\hat{a}^\dagger_\beta]\,=\,0\, .\ee
They act in an auxiliary Fock space ${\cal F}$ spanned by normalized
vectors
\be |n_1,n_2\rangle\ =\
\frac{(\hat{a}^\dagger_1)^{n_1}\,(\hat{a}^\dagger_2)^{n_2}}{
\sqrt{n_1!\,n_2!}}\ |0\rangle\,. \ee
Here, $|0\rangle\,\equiv\,|0,0\rangle$ denotes the normalized
vacuum state: $\hat{a}_1\,|0\rangle\ =\ \hat{a}_2\,|0\rangle\ =\
0$.

The noncommutative coordinates $\hat{x}_j$, $j\,=\,1,2,3$, in the 
space ${\bf\hat{R}}^3_0$ are given as
\be\label{sph3} \ \hat{x}_j\ =\
\lambda\,\hat{a}^+\,\sigma_j\,\hat{a}\ \equiv\
\lambda\,\sigma^j_{\alpha\beta}\,\hat{a}^\dagger_\alpha\,\hat{a}_\beta,\
j\,=\,1,2,3\,,\ee
where $\lambda$ is a universal length parameter. The coordinates
$\hat{x}_j$ satisfy rotationally invariant commutation rules:
\be\label{ncx} [\hat{x}_i,\hat{x}_j]\ =\
2i\,\lambda\,\varepsilon_{ijk}\,\hat{x}_k\,,\ \ \ \
[\hat{x}_i,\hat{\varrho }]\,=\,0\,,\ee
where $\hat{\varrho }\,=\,\lambda\,\hat{N}$, and $\hat{N}\,=\,
\hat{a}^+\,\hat{a}\,\equiv\,\hat{a}^\dagger_\alpha\,\hat{a}_\alpha$.

The operator that approximates the Euclidean distance from the
origin in an optimal way is $\hat{r}\,=\,\lambda\,(\hat{N} +
1)\,=\, \hat{\varrho} + \lambda $, and not $\hat{\varrho }$.
Namely, it holds $\hat{r}^2 - \hat{x}_j^2\,=\, \lambda^2$, whereas
$\hat{\varrho }^2 - \hat{x}_j^2\,=\, o(\lambda)$. In Section 3 we
give a strong argument supporting the exceptional role of
$\hat{r}$.

Let us consider the linear space of normal ordered polynomials containing
the same number of creation and annihilation operators:
\be\label{H_0} \hat\Psi\ =\ \sum\, C_{m_1 m_2 n_1 n_2}\,(\hat{a}^\dagger_1
)^{m_1}\,(\hat{a}^\dagger_2)^{m_2}\,(\hat{a}_1)^{n_1}\,(\hat{a}_2)^{n_2}\,,\ee
where the summation is finite over nonnegative integers satisfying
$m_1+m_2 \,=\,n_1+n_2$. In this space we define generators of
rotations $L_j$, $j\,=\,1,2,3$, as follows
\be\label{L0} L_j\,\hat\Psi\ =\ \frac{i}{2}\,[\hat{a}^+\,\sigma_j\,\hat{a},
\hat\Psi],\ \ j\,=\,1,2,3\,,\ee
obeying proper commutation relations
\be [L_i,L_j] \hat{\psi}\,\equiv\,(L_iL_j\,-\,L_jL_i)
\hat{\psi}\,=\, i\,\varepsilon_{ijk} L_k \hat{\psi}\ .\ee
With respect to the rotations (\ref{L0}) the doublet of
annihilation (creation) operators transforms as spinor (conjugated
spinor), whereas the triplet of NC coordinates as vector
\[ L_j\,\hat{a}_\alpha\,=\,-\,\frac{i}{2}\,\sigma^j_{\alpha\beta}
\,\hat{a}_\beta,\ \ L_j\,\hat{a}^\dagger_\alpha\ =\,\frac{i}{2}\
\sigma^j_{\beta\alpha}\ \hat{a}^\dagger_\beta\,,\ \ \
\ L_i\,\hat{x}_j\ =\ i\,\varepsilon_{ijk}\ \hat{x}_k\,.\]

The standard eigenfunctions $\hat{\psi }_{jm}$, $j =
0,1,2,\,\dots,\,$, $m = -j,\,\dots,\,+j$, sa\-tisfying
\be\label{Hjm1} L^2_i\,\hat{\psi }_{jm}\ =\ j(j+1)\,\hat{\psi }_{jm},\ \ \
L_3\,\hat{\psi }_{jm}\ =\ m\,\hat{\psi }_{jm}\ ,\ee
are given by the formula
\be\label{harm2} \hat{\psi }_{jm}\ =\ \lambda^j\ \sum_{(jm)}\
\frac{(\hat{a}^\dagger_1
)^{m_1}\,(\hat{a}^\dagger_2)^{m_2}}{m_1!\,m_2!}\ :R_j(\hat{\varrho
}):\ \frac{\hat{a}^{n_1}_1\,(-\hat{a}_2)^{n_2}}{n_1!\ n_2!} \ee
with the summation over all nonnegative integers satisfying
$m_1+m_2 \,=\,n_1+n_2\,=\,j$, $m_1-m_2-n_1+n_2\,=\,2 m$. Thus
$\hat{\psi }_{jm} = 0$ when restricted to the subspaces ${\cal
F}_N\, =\, \{ |n_1,n_2\rangle\,|\ n_1+n_2 = N\}$ with $N < j$. For
any fixed $:R_j(\hat{\varrho}):$ equation (\ref{harm2}) defines a
representation space for a unitary irreducible representation with
spin $j$.

The symbol $:R_j(\hat{\rho }):$ represents a normal ordered analytic function
in the operator $\hat{\varrho }$:
\be\label{harm3} :R_j(\hat{\varrho}):\ =\ \sum_k c^j_k\,:\hat{
\varrho}^k:\ =\ \sum_k c^j_k \lambda^k\,\frac{\hat{N}!}{
(\hat{N}-k)!}. \ee
The last equality follows from the equation
\be\label{nk} :\hat{N}^k:\,|n_1,n_2\rangle\ =\ \frac{N!}{(N-k)!}\
|n_1,n_2 \rangle,\ \ \ N\,=\,n_1 + n_2 \ee
(which can be proved by induction in $k$). Since, $:\hat{N}^k:\,
|n_1,n_2\rangle\,=\,0$ for $k\,>\,n_1 + n_2$, the summation in
(\ref{harm3}) is effectively restricted to $k \le N$ on any
subspace ${\cal F}_N$.

\section{Quantum mechanics in space $\boldsymbol{\hat{R}^3_0}$}

Let $\hat{\cal H}$ denote the Hilbert space generated by
functions (\ref{harm2}) with weighted Hilbert-Schmidt norm
\be\label{whs1} \|\hat\Psi \|^2\ =\ 4\pi\,\lambda^3\,\mbox{Tr}
[(\hat{N}+1)\,\hat{\Psi}^\dagger\,\hat{\Psi}]\ =\ 4\pi\,
\lambda^2\,\mbox{Tr}[\hat{r}\,\hat{\Psi}^\dagger\,\hat{\Psi}]\,, \
\ \hat{r}\,=\,\lambda\,(\hat{N}+1)\,.\ee

The rotationally invariant weight $w(\hat{r})\,=\,4\pi\,
\lambda^2\,\hat{r}$ is determined by the requirement that a ball
in ${\bf\hat{R}}^3_0$ with radius $r = \lambda (N+1)$ should
possess a standard volume in the limit $r\,\to\,\infty$. The
projector $\hat{P}_N$ on the subspace ${\cal F}_0
\oplus\,\dots\,\oplus {\cal F}_N$ corresponds to the
characteristic functions of a ball with the radius $r = \lambda
(N+1)$. Therefore, the volume of the ball in question in ${\bf
\hat{R}}^3_0$ is
\be V_r\ =\ 4\pi\,\lambda^3\,\mbox{Tr}[(\hat{N}+1)\,\hat{P}_N]\ =\
4\pi\,\lambda^3\,\sum_{n=0}^{N+1} (n+1)^2\ =\
\frac{4\pi}{3}\,r^3\,+\,o(\lambda)\,.\ee
Thus, the chosen weight $w(\hat{r})\,=\,4\pi\,\lambda^2\,\hat{r}$
possesses the desired property.

{\it Note}: The weighted trace $\mbox{Tr} [w(\hat{r})\,...\,]$
with $w(\hat{r})\,=\,4\pi\, \lambda^2\,\hat{r}$ at large distances
goes over to the usual volume integral $\int\,d^3\vec{x}\,...\,$. The 3D
noncommutative space proposed in \cite{Jab1} corresponds to the
choice $w(\hat{r})\,=\,const$ and at large distances does not
correspond to the flat space ${\bf {R}}^3_0$.

The generators of rotations $L_j$, $j\,=\,1,2,3$, are hermitian
(self-adjoint) operators in $\hat{\cal H}$, and consequently, the
two operators $\hat{\tilde\Psi}_{jm}$ and $\hat\Psi_{j'm'}$, with
arbitrary factors $:R_j(\hat{\varrho}):$ and
$:\tilde{R}_{j'}(\hat{\varrho}):$, are in $\hat{\cal H}$
orthogonal. It is sufficient to calculate  $\|\hat\Psi_{jm}
\|^2\,=\,\|\hat\Psi_{jj}\|^2$ (this equality follows from the
rotational invariance of the norm in question):
\be\label{psi2} \|\hat\Psi_{jm} \|^2\ =\ 4\pi \lambda^3\,\sum_{N=j}^\infty
\,\sum_{n=0}^N (N+1)\ \langle n,N-n|\,(\hat{N}+1)\,\hat{\Psi}_{jj}^\dagger
\,\hat{\Psi}_{jj}\,|n,N-n\rangle\,,\ee
We benefit from the fact that $\hat{\Psi}_{jj}$ has a simple form
\be\label{psijj} \hat{\Psi}_{jj}\ =\ \frac{\lambda^j}{(j!)^2}\
(\hat{a}^\dagger_1)^j \,:R_j(\hat{\varrho}):\,(-\hat{a}_2)^j \ee
The matrix element we need to calculate is
\[ \langle n,N-n|\ (\hat{a}^\dagger_2)^j \,:R_j(\hat{\varrho}):\,\hat{a}^j_1\
(\hat{a}^\dagger_1)^j\,:R_j(\hat{\varrho}):\,\hat{a}^j_2\ |n,N-n\rangle \]
\be\label{mel}  =\ \frac{(n+j)!(N-n)!}{n!\,(N-j-n)!}\ |{\cal R}_j(N-j)|^2,
\ \ee
where
\be {\cal R}_j(N)\,=\,\langle n,N-n|:R_j(\hat{\varrho}):
|n,N-n\rangle \ee
(the expression on the r.h.s. is $n$ - independent). Inserting (\ref{psijj}),
(\ref{mel}) into (\ref{psi2}) and using the identity
\[ \sum_{n=0}^N\ {{n+j}\choose j}\ {{N-n}\choose j}\ =\
{{N+j+1}\choose {2j+1}}\,,\]
we obtain
\be\label{Psi2} \|\hat\Psi_{jm} \|^2\ =\ \frac{4\pi
\lambda^{3+2j}}{(j!)^2}\ \sum_{N=0}^\infty\ (N+j+1)\
{{N+j+1}\choose {2j+1}}\ |{\cal R}_j(N)|^2 \,. \ee
This expression represents, up to an eventual normalization, the
square of a norm of the radial part of the operator wave
function.\skip0.5cm

Now we are ready to define the Coulomb problem Hamiltonian in the
noncommutative case. \skip0.5cm

{\it The kinetic term in} ${\bf\hat{R}}^3_0$. In the first we
define the NC analog of the Laplacian in ${\bf\hat{R}}^3_0$ as
follows:
\be\label{Lapl} \Delta_\lambda\,\hat{\Psi}\ =\ -\,\frac{1}{\lambda
\hat{r}}\ [\hat{a}^\dagger_\alpha,\,[\hat{a}_\alpha
,\,\hat{\Psi}]]\ =\ -\,\frac{1}{\lambda^2 (\hat{N}+1)}\
[\hat{a}^\dagger_\alpha ,\,[\hat{a}_\alpha ,\,\hat{\Psi}]]\,.\ee
This choice is motivated by the following facts:

$\bullet $ A double commutator is an analog of a second order
differential operator;

$\bullet $ The factor  $\hat{r}^{-1}$ guarantees that the operator
$\Delta_\lambda$ is hermitian (self-adjoint) in $\hat{\cal H}$,
and finally,

$\bullet $ The factor $\lambda^{-1}$, or $\lambda^{-2}$
respectively, guarantees the correct physical dimension of
$\Delta_\lambda$ and its non-trivial commutative limit.\vskip0.5cm

Calculating the action of (\ref{kinH}) on  $\hat{\psi }_{jm}$
given in (\ref{harm2}) we can check whether the postulate
(\ref{kinH}) is a reasonable choice. The corresponding formula is
derived in Appendix A:
\[ -\,[\hat{a}^\dagger_\alpha ,\,[\hat{a}_\alpha
,\,\hat{\Psi}]]\ =\ \lambda^j\ \sum_{(jm)}\
\frac{(\hat{a}^\dagger_1)^{m_1}\,
(\hat{a}^\dagger_2)^{m_2}}{m_1!\, m_2!}\]
\be\label{rrov0} \times\ \ :[ \hat{\varrho }\,R''(\hat{\varrho
})\,+\,2(j+1)\,R'(\hat{\varrho })]:\
\frac{\hat{a}^{n_1}_1\,(-\hat{a}_2)^{n_2} }{n_1!\ n_2!}\,.\ee
Here the symbols $R'(\hat{\varrho })$ and $R''(\hat{\varrho })$
are {\it defined} as:
\beqa\nn  R(\hat{\rho })\,=\,\sum_{k=0}^\infty\,c^j_k\,\hat{\rho
}^k\ \ \ \Rightarrow \ \ \  R'(\hat{\varrho
})\,&=&\,\sum_{k=1}^\infty\, k\,c^j_k\,\hat{\varrho }^{k-1} ,\\
\label{deriv} R''(\hat{\varrho })\,&=&\,\sum_{k=2}^\infty\,
k(k-1)\,c^j_k\,\hat{\varrho }^{k-2}\,.\eeqa
Thus, the prime corresponds exactly to the usual derivative
$\partial_{\hat{\varrho}}$. We see that the angular dependence 
in (\ref{rrov0}) remains untouched, since $\Delta_\lambda$ is 
rotation invariant. In the commutative limit $\lambda\,\rightarrow
\,0$ formally $\hat{\varrho}\,\rightarrow\,r$, and we see that 
(\ref{rrov0}) guarantees that $\Delta_\lambda$ reduces just to 
the standard Laplacian.

Based on that, we postulate the kinetic term of the Hamiltonian as
follows
\be\label{kinH} H_0\ =\ -\,\frac{\hbar^2}{2m}\,\Delta_\lambda\,
\hat{\Psi}\ =\ \frac{\hbar^2}{2m \lambda \hat{r}}\
[\hat{a}^\dagger_\alpha,\,[\hat{a}_\alpha ,\,\hat{\Psi}]]\,.\ee

\vskip0.5cm

{\it The potential term in} ${\bf\hat{R}}^3_0$ is defined by left
multiplication  of the operator wave function by potential
$\hat{V} = V(\hat{x})$: $\hat{\Psi }\ \mapsto\ \hat{V}\,\hat{\Psi
}$. The potential is central if $\hat{V}\,=\,V(\hat{N})$, or
equivalently, $\hat{V}\,=\,V(\hat{r})$.

\section{The Coulomb problem in $\boldsymbol{\hat{R}^3_0}$}

In the commutative case the Coulomb potential is a radial solution
of the equation (\ref{coul0}) finite at infinity. Due to our
choice of the noncommutative Laplacian $\Delta_\lambda$ the
equivalent equation in ${\bf\hat{R}}^3_0$ is
\[ [\,\hat{a}^\dagger_\alpha ,\,[\,\hat{a}_\alpha\,,V(\hat{N})\,]\,]\
=\ 0\,. \]
It can be rewritten as a simple recurrent relation
\be\label{rec} (\hat{N}+2)\,V(\hat{N}+1)\,-\,
(\hat{N}+1)\,V(\hat{N})]\ =\ (\hat{N}+1)\,[V(\hat{N})\,-\,
\hat{N}\,V(\hat{N}-1)\,.\ee
Putting
\[ (\hat{M}+1)\,V(\hat{M})\,-\,\hat{M}\,V(\hat{M}-1)\ =\ q_0
\,,\ \ \ V(0)\ =\ q_0\,-\,\frac{q}{\lambda}\,,\]
and summing up the first equation over $M\,=\,1,\,\dots\,N$, we
obtain the general solution:
\be\label{nccoul} V(\hat{N})\ =\ -\,\frac{q}{\lambda\,
(\hat{N}+1)}\ +\ q_0\ =\ -\,\frac{q}{\hat{r}}\ +\ q_0\,,\ee
where, $q$ and $q_0$ are arbitrary constants ($\lambda$ is
introduced for convenience). For $H$-atom $q = e^2$ and we put
$q_0 = 0$. We see that the dependence $\hat{r}^{-1}$ of the NC
Coulomb potential is inevitable. \vskip0.5cm

Thus, the noncommutative analog of the Schr\"odinger equation with
the Coulomb potential in ${\bf\hat{R}}^3_0$ is
\be\label{ncsch1} \frac{\hbar^2}{2m\lambda \hat{r}}\,
[\hat{a}^\dagger_\alpha , [\hat{a}_\alpha ,\hat{\Psi}]] -
\frac{q}{\hat{r}}\, \hat{\Psi} = E\,\hat{\Psi}\ \ \
\Leftrightarrow\ \ \ \frac{1}{\lambda}\,[\hat{a}^\dagger_\alpha,
[\hat{a}_\alpha ,\hat{\Psi}]] - \alpha\,\hat{\Psi} =
-\,\kappa^2\,\hat{r} \hat{\Psi}\,,\ee
where $\alpha = 2me^2/\hbar^2$ and $\kappa = \sqrt{-2m E} /\hbar$.
The nocommutative corrections comming from $\hat{r} \hat{\Psi}$ 
are calculated in Appendix A:
\be\label{appen} \hat{r}\,\hat{\Psi}_{jm}\ =\  \sum_{(jm)}\ 
\frac{(\hat{a}^\dagger_1)^{m_1}\,(\hat{a}^\dagger_2)^{m_2}}{m_1!\, m_2!}
\ :[(\hat{\varrho }+\lambda j+\lambda )\,\hat{R}_j\,+\,\lambda
\,\hat{\varrho }\,\hat{R}_j^\prime]:\ 
\frac{\hat{a}^{n_1}_1\,(-\hat{a}_2)^{n_2} }{n_1!\ n_2!} \ .\ee
where $\hat{R}_j \equiv R_j(\hat{\varrho })$ and similarly for
derivatives, e.g., $\hat{R}_j^\prime \equiv R_j^\prime(
\hat{\varrho})$. The angular dependence is again untouched as the 
multiplication by $ \hat{r}$ represents a rotation invariant operator.

Inserting (\ref{rrov0}) and (\ref{appen}) into (\ref{ncsch1}) we obtain 
the NC analog of radial Schr\"odinger equation:
\be\label{ncradsch} :[ \hat{\varrho }\,\hat{R}_j^{\prime\prime}
\,+\, 2(j+1)\,\hat{R}_j^\prime\ +\ \alpha\, \hat{R}_j]:\ =\
\kappa^2\,:[\hat{\varrho }\hat{R}_j\,+\,\lambda ((j+1)
\,\hat{R}_j\,+\,\hat{\varrho }\,\hat{R}_j^\prime)]:\,.\ee
The term proportional to $\lambda $ represents the NC correction.
For $\lambda\,\rightarrow \,0 $ this NC radial Schr\"odinger
equation reduces to the standard radial Schr\"odinger equation.

We associate the following ordinary differential equation to the
mentioned operator radial Schr\"odinger equation (\ref{ncradsch}):
\be\label{ncradsch1} \varrho \,{\cal R}_j^{\prime\prime}\,+\,
2(j+1)\,{\cal R}_j^\prime \,+\,\alpha\, {\cal R}_j\ =\
\kappa^2\,[\varrho\,{\cal R}_j\,+\,\lambda ((j+1)\,{\cal
R}_j\,+\,\varrho\,{\cal R}_j^\prime)]\,.\ee
If the function ${\cal R}_j\,=\,{\cal R}_j(\varrho)\,=\,\sum_k
c^j_k\,\varrho^k$ solves the associated ordinary differential
equation (\ref{ncradsch1}), then
\be\label{hatR} \hat{R}_j\ =\ :{\cal R}_j(\hat{\varrho}):\ =\
\sum_k c^j_k\,:\hat{\varrho}^k:\ =\ \sum_k c^j_k
\lambda^k\,\frac{\hat{N}!}{(\hat{N}-k)!}\,. \ee
solves the operator radial equation (\ref{ncradsch}). Moreover,
{\it the operator function $\hat{R}_j= \ :{\cal R}_j(\hat{\varrho}):$
possesses a finite norm in $\hat{\cal H}$ provided the function
${\cal R}_j\,=\,{\cal R}_j(\varrho)$ has finite norm in ${\cal
H}$} (since the norm (\ref{whs1}) asymptotically reduces to the
usual QM norm).

The solution ${\cal R}_j$ of the associated radial Schr\"odinger
equation (\ref{ncradsch1}) is given  similarly as in the standard
Coulomb problem in (\ref{rad0}), but with particularly scaled
$\varrho$ dependence in the exponent and in the argument of the
confluent hypergeometric function:
\be\label{sol2} {\cal R}_j(\varrho )\ =\ \,e^{-b\,\kappa \varrho
}\ F\left(j+1 - \frac{\alpha }{2d\,\kappa},\,2j+2\,;2\varrho
\kappa\,d \right)\,.\ee
The dimensionless quantities $b$ and $d$ given as (see Appendix B)
\be\label{nccor} b\ =\ \,\sqrt{1 + \eta^2}\,-\,\eta\,,\ \ \ d\ =\
\sqrt{1 + \eta^2}\,,\ \ \eta\,=\,\frac{1}{2}\,\lambda \kappa \ee
specify the NC corrections to the usual radial function in
(\ref{rad0}). They enter $b$ and $d$ via the parameter
$\eta\,=\,\lambda \kappa /2$. So they vanish not only in the
commutative limit $\lambda\,\to\,0$ , but also for $\kappa\,\to\,0$.

We shall restrict ourselves to the determination of the bound
states spectra with $E < 0$ and $\kappa > 0$. In this case ${\cal
R}_j(\varrho )$ should be normalizable. This is ensured if the
first argument of the confluent hypergeometric function is zero or
negative integer, what determines the discrete energy eigenvalues
(remember that $d$ is $\kappa$-dependent, see (\ref{nccor})):
\[
\frac{\alpha}{2d_n\kappa_n}\,=\,n\,=\,j+1,\,j+2,\,\dots\ \ \
\Rightarrow \]
\be\label{ener1} E^\lambda_n\,=\,-\frac{\hbar^2}{2m}\,\kappa^2_n
\,=\,-\frac{m e^4 }{2\hbar^2 n^2}\
\frac{2}{1+\sqrt{1+\lambda^2/a^2_0 n^2}}\,, \ee
where $a_0 = 2/\alpha = \hbar^2/m e^2\, =\, 5.29\,\times\,10^{-11}
m$ is the Bohr radius. The first factor in $E^\lambda_n$ is just
the standard bound state energy of the Coulomb problem, whereas
the second one represents the noncommutative correction. The NC
corrections in the limit $\lambda/n \to 0$, i.e., in the
commutative limit $\lambda \to 0$, or for fixed $\lambda$, in the
quasi-classical limit $n \to \infty$ for highly excited states.

The solutions $\hat{\Psi }^\lambda_{njm}$ of the operator equation
(\ref{ncradsch}) corresponding to the energy $E^\lambda_n$ is
\be\label{wfn} \hat{\Psi }^\lambda_{njm}\ =\ N^{\lambda n}_{jm}\
\sum_{(jm)}\ \frac{(\hat{a}^\dagger_1)^{m_1}\,(\hat{a}^\dagger_2
)^{m_2}}{m_1!\, m_2!}\ :{\cal R}_{nj}(\hat{\varrho }):\
\frac{\hat{a}^{n_1}_1\,(-\hat{a}_2)^{n_2}}{n_1!\ n_2!}\,, \ee
where $N^{\lambda n}_{jm}$ denotes the normalization factor and
\be\label{Rnjm} {\cal R}_{nj}(\hat{\varrho })\ =\
e^{-b_n\,\kappa_n \hat{\varrho}}\
F\left(j+1-n,\,2j+2,\,2\hat{\varrho} \kappa_n\,d_n \right)\, .\ee
Here, the parameters $b_n$ a $d_n$ are given by (\ref{nccor}) with
$\kappa_n \,=\,\sqrt{-2 m E^\lambda_n}/\hbar$. From the equation
(\ref{nk}) it follows directly
\be\label{nk1} :e^{-a\hat{\varrho}}\,\hat{\varrho}^k:\ =\ (1 -
a\lambda)^{\hat{N}}\ \frac{\lambda^k}{(1 - a \lambda)^k}\,
\frac{\hat{N}!}{(\hat{N}-k)!}\ =\ (1 - a \lambda)^{\hat{\varrho}/
\lambda}\ \frac{:\hat{\varrho}^k:}{(1 - \lambda a)^k}\,.\ee
This allows as to express the normal ordered form of the radial
part of operator wave function:
\[ :{\cal R}_{nj}(\hat{\varrho }):\,=\,(1 -
b_n\,\kappa_n\lambda)^{\hat{\varrho}/\lambda}\,
:F\left(j+1-n,\,2j+2,\,\frac{2\hat{\varrho} \kappa_n d_n}{1 -
b_n\,\kappa_n\lambda} \right):\]
\be\label{wfn1} =\ (1 - b_n\,\kappa_n\lambda)^{\hat{N}}\,
\sum_{k=0}^{n-j-1} c^k_{nj}\,\frac{ (-2\lambda\kappa_n d_n)^k}{(1
- b_n\,\kappa_n\lambda)^k}\,{{\hat{N}}\choose k}\ , \ee
with the coefficients $c^k_{nj}$ given in (\ref{deghf}). Inserting
\be {\cal R}_{nj}(N)\ =\ N^{\lambda n}_{jm}\ (1 -
b_n\,\kappa_n\lambda)^N\, \sum_{k=0}^{n-j-1} c^k_{nj}\,\frac{
(-2\lambda\kappa_n d_n)^k}{(1 -
b_n\,\kappa_n\lambda)^k}\,{N\choose k}\ , \ee
into (\ref{Psi2}) the normalization constant $N^{\lambda n}_{jm}$
can be determined (we skip its calculation). In commutative limit
the factor in front of the sum gives the usual exponential damping
factor in (\ref{rad0}). However, as the argument of the polynomial
becomes scaled due to NC corrections, the separation of the
asymptotic factor from the polynomial part is not perfect.

\section{Conclusions}

We carefully defined the NC rotationally invariant analog of the
QM configuration space and the Hilbert space of operator wave
functions in NC configuration space. The central point of our
construction was the definition of $\Delta_\lambda$ the NC analog
of Laplacian, supplemented by a consequent definition of the
weighted Hilbert-Schmidt norm and a definition of the Coulomb
potential satisfying NC Poisson equation.

With this input this Hilbert space we introduced the NC analog of
$H$-atom Hamiltonian and explicitly determined the bound-state
energies $E^\lambda_n$ and corresponding eigenstates
$\hat{\psi}^\lambda_{njm}$ (see equations (\ref{ener1}),
(\ref{wfn}) and (\ref{Rnjm})).

We found that the discrete parameters $n,\,j,\,m$ have the same
meaning and range as in the standard (commutative) Coulomb
problem, and moreover, the bound-state energies and eigenstates
possess a smooth commutative limit $\lambda \to 0$. This paper
does not deal with the case of the scattering in the NC
configuration space - this will be discussed elsewhere.

The noncommutativity parameter $\lambda$ is not fixed within our
model. However, it can be estimated by some other physical
requirement. For example, one can postulate, as was done in early
days of modern physics, that the rest energy $mc^2$ of electron is
equal to the electrostatic energy of its Coulomb field. In
${\bf\hat{R}}^3_0$ this means:
\be\label{ncradius} mc^2\ =\ \frac{4\pi
\lambda^3}{8\pi}\,\mbox{Tr}\,[ (\hat{N}+1)\, \hat{E}^2_j ]\,\ee
where
\be \hat{E}_j\ =\ \frac{e^2}{\lambda^3}\,\frac{1}{\hat{N}
(\hat{N}+1)(\hat{N}+2)}\,\hat{x}_j\,,\ee
is the NC electric field strength corresponding to NC Coulomb
potential $\hat{\Phi}$ which was discussed in Section 4 (the
details will be published, see \cite{KP}).

We stress that in the NC case the electrostatic energy of
electron, determined by the trace in (\ref{ncradius}), is finite
(no cut-off at short distance is needed). A straightforward
calculation of the trace in (\ref{ncradius}) gives the relation:
\be mc^2\ =\ \frac{3}{8}\,\frac{e^2}{\lambda}\ \ \ \Rightarrow\ \
\ \lambda\ =\ \frac{3}{8}\,\frac{e^2}{mc^2}\,\equiv\,\lambda_0
\,.\ee
This $\lambda_0$ is fraction of the classical radius of electron
$r_0 = e^2/mc^2$: $\lambda_0 = 1.06\,\times\,10^{-15}\,m = 1.06\,
fm$ (the coincidence with the proton radius is purely accidental).

The NC corrections to the $H$-atom energy levels given in
(\ref{ener1}) are of of order $(\lambda_0/a_0)\, =\, (9/64)\,
\alpha^2_0\,\approx\, 4\,\times\,10^{-11}$ (here $\alpha_0 \approx
1/137$ is fine structure constant). Such tiny corrections to
energy levels are beyond any experimental evidence. Moreover, at
$\lambda_0 \approx 1 \,fm$ relativistic and QFT effects become
essential.

Our investigation indicates that the noncommutativity of the
configuration space is fully consistent with the general QM
axioms, at least for the  $H$-atom bound states. However, a more
detailed analysis of the Coulomb problem in ${\bf\hat{R}}^3_0$
would be a desirable dealing, e.g, with the following aspects:

$\bullet$ Coulomb scattering problem, dyon problem (electron in
the electric point charge and magnetic monopole field), Pauli
$H$-atom (non-relativistic spin);

$\bullet$ Coulomb problem in ${\bf\hat{R}}^3_0$ and its dynamical
symmetry, QM supersymmetry and integrability of the Coulomb
system.

Besides non-relativistic $H$-atom, there are other systems that
would be interesting to investigate within NC configuration space
${\bf\hat{R}}^3_0$=, e.g., Dirac $H$-atom (relativistic
invariance?), non-Abelian monopoles, or spherical black-holes.

\vskip0.5cm

{\bf Acknowledgements}: The authors would like to thank to M.
Chaichian, M. M. Sheikh-Jabbari and A. Tureanu for valuable
comments. This work was supported by project VEGA 1/100809/1.

\vskip1cm 

{\bf\Large Appendix A}\vskip0.5cm

Here we prove two formulas, (\ref{rrov0}) and (\ref{appen}), we need for the
calculation of the NC Coulomb Hamiltonian. Below we skip indices $j$ and
$m$.

a. Let us begin with the (\ref{rrov0}):
\beqa\nn [\hat{a}^\dagger_\alpha ,[\hat{a}_\alpha
,\hat{\Psi}_{jm}]] &=& \lambda^j [\hat{a}^\dagger_\alpha
,\,[\hat{a}_\alpha ,\sum_{(jm)}\frac{(\hat{a}^\dagger_1
)^{m_1}\,(\hat{a}^\dagger_2)^{m_2}}{m_1!\,m_2!}\ :\hat{R}:\
\frac{\hat{a}^{n_1}_1\,(-\hat{a}_2)^{n_2}}{n_1!\ n_2!}]]\\
\nn &=&\ \lambda^j \sum_{(jm)}[\hat{a}_\alpha
,\frac{(\hat{a}^\dagger_1 )^{m_1}
(\hat{a}^\dagger_2)^{m_2}}{m_1!\,m_2!}]\,
[\hat{a}^\dagger_\alpha,:\hat{R}:]\,
\frac{\hat{a}^{n_1}_1(-\hat{a}_2)^{n_2}}{n_1!\ n_2!} \\
\nn &+&\ \lambda^j \sum_{(jm)}[\hat{a}_\alpha
,\,\frac{(\hat{a}^\dagger_1
)^{m_1}\,(\hat{a}^\dagger_2)^{m_2}}{m_1!\,m_2!}]\,:\hat{R}:\,
[\hat{a}^\dagger_\alpha,\,\frac{\hat{a}^{n_1}_1
(-\hat{a}_2)^{n_2}}{n_1!\ n_2!}] \\
\nn  &+& \lambda^j \sum_{(jm)}\frac{(\hat{a}^\dagger_1
)^{m_1}\,(\hat{a}^\dagger_2)^{m_2}}{m_1!\,m_2!}\,
[\hat{a}^\dagger_\alpha,\,[\hat{a}_\alpha ,:\hat{R}:]]\,
\frac{\hat{a}^{n_1}_1\,(-\hat{a}_2)^{n_2}}{n_1!\
n_2!} \\
\label{A1} &+& \lambda^j \sum_{(jm)}\frac{(\hat{a}^\dagger_1
)^{m_1}\,(\hat{a}^\dagger_2)^{m_2}}{m_1!\ m_2!}\ [\hat{a}_\alpha
,:\hat{R}:]\, [\hat{a}^\dagger_\alpha,\frac{\hat{a}^{n_1}_1
\,(-\hat{a}_2)^{n_2}}{n_1!\ n_2!}] ,  \eeqa
where $\hat{R}\,=\,\sum_{k=0}^\infty\,\tilde{c}^j_k\,\hat{N}^k$,
i.e., $\tilde{c}^j_k = \lambda^j\,c^j_k$. Now we shall use the
following commutation relations
\beqa\nn [\hat{a}^\dagger_\alpha,:\hat{N}^k:]\,=- \,k\,
\hat{a}^\dagger_\alpha\, :\hat{N}^{k-1}:\ \ &\Rightarrow& \ \
[\hat{a}^\dagger_\alpha,:\hat{R}:]\,=\, -\hat{a}^\dagger_\alpha\,
:\partial_{\hat{N}}\hat{R}:\,,\\
\label{A2 } [\hat{a}_\alpha,:\hat{N}^k:]\,=\, k\,
:\hat{N}^{k-1}:\,\hat{a}_\alpha\ \ &\Rightarrow& \ \
[\hat{a}_\alpha,:\hat{R}:]\,=\,:\partial_{\hat{N}}\hat{R}:
\,\hat{a}_\alpha \,,\eeqa
where $\partial_{\hat{N}}$ denotes the derivatives with respect to
$\hat{N}$: $\partial_{\hat{N}}\hat{R}\,=\,\sum_{k=1}^\infty
k\,\tilde{c }^j_k\, \hat{N}^{k-1}$.

It is easy to see that the second line in (\ref{A1}) vanish, and
the the first and third line give the same contribution
\be\label{A3 }  \sum_{(jm)}\,\frac{(\hat{a}^\dagger_1
)^{m_1}\,(\hat{a}^\dagger_2)^{m_2}}{m_1!\,m_2!}\
(-j\,:\partial_{\hat{N}}\hat{R}:)\
\frac{\hat{a}^{n_1}_1\,(-\hat{a}_2)^{n_2}}{n_1!\ n_2!}\,.\ee
From (\ref{A2 }) the double commutator
$[\hat{a}^\dagger_\alpha,[\hat{a}_\alpha,:\hat{R}:]]$ follows
directly, and this gives the value of the third line in (\ref{A1})
\be\label{A4 } \sum_{(jm)}\,\frac{(\hat{a}^\dagger_1 )^{m_1}\,
(\hat{a}^\dagger_2)^{m_2}}{m_1!\,m_2!}\ (-:\hat{N}\,
\partial^2_{\hat{N}} \hat{R}:\,+\,2\,:\partial_{\hat{N}}\hat{R}:)\
\frac{\hat{a}^{n_1}_1\,(-\hat{a}_2)^{n_2}}{n_1!\ n_2!}\,.\ee
Introducing parameter $\lambda$ and switching to the derivatives with
respect to $\hat{\varrho}$ the last two equations yields (\ref{rrov0}).

b. Proof of (\ref{appen}) is straightforward. From equation (\ref{nk}) 
it follows easily
\be\label{A5} \hat{N}\,:\hat{N}^k:\ =\ :\hat{N}^{k+1}:\,+\,k\,:
\hat{N}^k:\ \ \ \Rightarrow\ \ \ \hat{N}\,:\hat{R}:\ =\
:\hat{N}\,\hat{R}:\,+\,:\hat{N}\,\partial_{\hat{N}}\hat{R}:\,.\ee
This relation gives directly
\[ (\hat{N} +1)\,\sum_{(jm)}\,\frac{(\hat{a}^\dagger_1
)^{m_1}\,(\hat{a}^\dagger_2)^{m_2}}{m_1!\,m_2!}\ :\hat{R}:\
\frac{\hat{a}^{n_1}_1\,(-\hat{a}_2)^{n_2}}{n_1!\ n_2!}\]
\beqa\nn &=& \ \sum_{(jm)}\,\dots\ [(\hat{N} +j +1)\,:\hat{R}:]\
\dots \\
\label{A6} &=& \ \sum_{(jm)}\,\dots\ :[(\hat{N} +j
+1)\,\hat{R}\,+\,\hat{N}\,\partial_{\hat{N}}\hat{R}]:\
\dots\,,\eeqa
where we have replaced  both untouched factors containing
annihilation and creation operators by dots. Introducing parameter
$\lambda$ again and switching to the derivatives with respect to
$\hat{\varrho}$ we recover (\ref{appen}).

\vskip1cm {\bf\Large Appendix B} \vskip0.5cm

It is known that the solution $R = R(\varrho)$ of equation
\be \varrho\,R^{\prime\prime}\,+\,(a_1 \varrho\,+\,b_1)R'\,+\,(a_2
\varrho\,+\,b_2)R\ =\ 0 \ee
can be expressed in terms of a confluent hypergeometric function
\[ F(a,c;x)\ =\ 1\,+\,\frac{a}{c} \frac{x}{1!}\,+\,\frac{a(a+1)}{c(c+1)}
\frac{x^2}{2!}\,+\ \dots\ .\]
The formula, given e.g. in \cite{Bat}, reads
\be R\ =\ e^{\frac{1}{2}(D-a_1)\varrho}\,F(a,c;-D\varrho)\,,\ee
where $D$ is determined by $D^2\,=\,a^2_1 - 4 a_2$ and
\be a\ =\ \frac{1}{D}\,\left[\frac{1}{2}\,(D-a_1)\,b_1\,+\,b_2
\right]\,,\ \ \ c\ =\ b_1\,. \ee

In our case
\be a_1\,=\,- \lambda\,\kappa^2,\ \ b_1\,=\,2 j\,+\,2,\ \ a_2\,=\,
- \kappa^2,\ \ b_2\,=\,\alpha\,-\,(j+1)\,\lambda\,\kappa^2\,,\ee
what gives
\be R(\varrho)\ =\ e^{\frac{1}{2}(D+\lambda\kappa^2)\rho}\
F\left(j+1 +\frac{\alpha}{D},\,2j+2 ;\,-D\,\varrho\right)\,.\ee
There are two solutions $R_\pm(\varrho)$ depending on the sign of
$D\,=\,\pm\,2\kappa\, \sqrt{1+\eta^2}$, $\eta\,=\,\frac{1}{2}\,
\lambda\,\kappa$. However, $R_+ (\varrho)\,=\,R_- (\varrho)$ due
the Kummer relation
\[ F(a,c;x)\,=\,e^x\,F(c-a,c;-x)\,.\]
We took the solution (\ref{Rnjm}) of the associated radial NC
Schr\" odinger equation (\ref{sol2}) with negative $D$ as it is
more convenient for the determination of bound states.

\end{document}